\documentclass[aps]{revtex4-1}
\usepackage{amsfonts,amssymb,wrapfig}
\def\beq{\begin{eqnarray}}
\def\eeq{\end{eqnarray}}
\def\mb#1{\mathbb{#1}}
\begin{document}
\title{Black hole state counting in loop quantum gravity
}
\author{A. Ghosh}\email{amit.ghosh@saha.ac.in}
\author{P. Mitra}\email{parthasarathi.mitra@saha.ac.in}
\affiliation{Saha Institute of Nuclear Physics,
1/AF Bidhannagar, Calcutta 700064}
%\date{}
\begin{abstract}
The two ways of counting microscopic
states of black holes in the U(1) formulation of
loop quantum gravity, one counting all
allowed spin network labels $j,m$ and the other only $m$ labels,
are discussed in some detail. The constraints on $m$ are 
clarified and the map between the flux quantum numbers and $m$ discussed.
Configurations with $|m|=j$, which are sometimes sought after,
are shown to be important only when large areas are involved. The discussion is
extended to the SU(2) formulation.
\end{abstract}
%\pacs{04.70.Dy, 04.60.Pp}
\maketitle
%\bigskip
%\bigskip
%\newpage
%%%%%%%%%%%%%%%%%%%%%%

%\section
{\it Introduction:} 
Loop quantum gravity has yielded a detailed prescription for identifying
microscopic quantum states corresponding to an isolated horizon
\cite{rov,ashck,ash}. The horizon quantum states arise
when the cross sections of the horizon are punctured by spin networks 
that live in the bulk. It can be shown that the spin quantum numbers
$j,m$, which characterize the spin network, can also be used to label the
quantum states of the horizon. The states counted are the ones
that are consistent with a fixed area of the cross section
and the boundary conditions imposed on the horizon.
An improved estimation of the number of states was carried out in \cite{meissner}
counting only configurations with distinct $m$-labels -- see also \cite {gm2}.
In an alternative scheme \cite{gm}, the $j$-labels are also
recognized as characterizing black hole microstates.
In counting $m$-labels as in \cite{meissner}, 
an additional $|m|=j$ projection may come into play \cite{dom}. 
In this work we propose to elucidate all restrictions on these $m$ quantum 
numbers and the $|m|=j$ prescription.
It should be clarified that although the horizon area is taken to be more 
or less fixed,
the complications which arise from fixing it with precision \cite{gm3} are
avoided here.

It may be noted that the above calculations were done in the framework of
a $U(1)$ Chern-Simons theory of isolated horizons. Instead, the counting of 
states can also be carried out in an unbroken $SU(2)$
Chern-Simons theory ({\it cf.} \cite{pm}),
an isolated horizon formulation for which is possible \cite{perez}.
We shall comment on the differences which arise in this formulation.

In the $m$-counting scheme -- originally proposed by \cite{ash} 
and used in \cite{meissner,gm2} 
-- one counts only surface states labelled by the flux quantum
numbers associated with punctures on the horizon. There is a 
map between the flux quantum numbers and $m$, so that 
one counts only those $m$-states that obey an area bound. 
On the other hand, $j$ quantum numbers label area eigenstates, which belong 
to the {\it bulk} Hilbert space. In $m$-counting,
states carrying different $j$ quantum numbers but the same $m$ quantum number 
are considered equivalent \cite{dom,gm2}, while in $(j,m)$-counting, such states are 
{\em not} identified. In $(j,m)$-counting the implementation of the area constraint 
is rather straightforward, while in $m$-counting there can only be a bound 
$|m|\leqslant j$, so that the area constraint has to be implemented only through 
some inequality. In the following we show that a one-to-one map from the 
flux quantum numbers to $m$ exists for both counting schemes. 

%\section
{\it Restrictions on $m$:} 
We start by showing that a proof of \cite{dom} can be 
tailored to suit the $(j,m)$-scheme \cite{gm}, from which one can also extract 
the $m$-scheme \cite{gm2}, that gives us a one-to-one map from
the flux quantum numbers to $m$. Before we go into the technical details, let us
choose units such that $4\pi\gamma\ell_P^2=1$, where $\gamma$ is the
Barbero-Immirzi parameter involved in the quantization and $\ell_P$ the
Planck length. In these units, counting is possible only for integral classical
areas
\beq A_{class}=k, \eeq
where $k$ is a positive integer labelling the Chern-Simons theory on the
horizon. This poses no serious problem in the semiclassical
limit, since assuming $\gamma\sim o(1)$, the pre-factor $4\pi\ell_P^2\sim
10^{-69}\,{\rm m}^2$, which is an extremely small number. So the area intervals
$\Delta A$ for which no counting is possible are also
extremely small, which essentially gives a continuum of classical areas.
The area eigenvalue for a given configuration of spins is
\beq A_{N[j,m]}=2\sum_{j,m}N[j,m]\sqrt{j(j+1)}\label{areacon}\eeq
where $N[j,m]$ is the number of punctures on the horizon carrying spin quantum numbers $(j,m)$.

The {\it horizon} states are labelled by the flux quantum numbers $b$
on the punctures; these are elements of $\mb Z_k$ obeying the restriction
\beq \sum_bN_bb=0\;\mbox{mod}\;k\label{fluxc}\eeq
where $N_b$ is the number of punctures carrying the flux quantum number $b$.
The quantum isolated horizon boundary conditions imply that the flux quantum
number $b$ associated with each puncture must be related to the spin projection 
quantum number $m$ associated with that puncture as
\beq b=-2m \;{\rm mod~} k.\label{boundc}\eeq
Then by (\ref{fluxc}), the spin projections obey the constraint
\beq \sum_{j,m}N[j,m]m=\frac{nk}{2}\label{five}\eeq
where $n\in\mb N$ is some integer. This constraint will now be sharpened, 
as in \cite{dom}, but in our analysis we shall keep the $j$ instead of
replacing them through inequalities by $m$ so that the choice between
the two kinds of counting is left open.

The area eigenvalues of interest are taken in a range
\beq k-\epsilon\leqslant A_{N[j,m]}\leqslant k+\epsilon\label{arear}\eeq
where $\epsilon<k$ is a positive number, a macroscopic parameter 
that should be independent of the microscopic configurations which are 
summed over in calculating the total number of microstates. Now
\beq A_{N[j,m]}&\geqslant& 2\sum_{j,m}N[j,m]\sqrt{|m|(|m|+1)}\quad\mbox{since}\;j \geqslant|m|\nonumber\\
&\geqslant& 2\sum_{j,m}N[j,m]|m|+\sum_{j,m}N[j,m](\sqrt 3-1)
\label{areac}\eeq
where in the last step we have made use of the fact that the quantity $[|m|(|m|+1)]^{1/2}-|m|$
increases monotonically with $|m|$ and $(\sqrt 3-1)/2$ is its lowest value, 
obtained from $m=1/2$, $m=0$  being unphysical because of the boundary 
condition (\ref{boundc}) which says that for $m=0$ the flux quantum number 
$b=0$ mod $k$ and such punctures are 
invisible in Chern-Simons theory. Now by (\ref{five}) the series
\beq \sum_{j,m}N[j,m]|m|\geqslant|\sum_{j,m}N[j,m]m|=\frac{|n|k}{2}.\label{sumc}\eeq
Thus since $k+\epsilon\geqslant A_{N[j,m]}$, from (\ref{areac}) and (\ref{sumc}) we get
\beq k+\epsilon\geqslant|n|k+\sum_{j,m}N[j,m](\sqrt 3-1).\label{ineq}\eeq
So for all nonzero values of $|n|$ either we cannot choose $\epsilon$ 
irrespective of the microscopic configurations (for $|n|=1$) or 
$\epsilon$ becomes larger than $k$ (for $|n|>1$) since $\sum_{j,m}N[j,m]>1$. 
The only allowed value is $n=0$. So the {\em spin-projection constraint} 
takes a more restricted form
\beq \sum_{j,m}N[j,m]m=0.\label{spinc}\eeq

The sum $\sum N[j,m](\sqrt 3-1)=N(\sqrt 3-1)$, where $N$ is the total number 
of punctures, is a large number $\sim o(k)$ for a large black hole. Hence 
$\epsilon<N(\sqrt 3-1)$. From (\ref{areac}) we get 
$2\sum N[j,m]|m|\leqslant k+[\epsilon-N(\sqrt 3-1)]<k$. 
So for a large black hole ($k\gg 1$) we get a further restriction
\beq \sum_{j,m}N[j,m]|m|<\frac{k}{2}.\label{newc}\eeq
If one is interested in small values of $k$, {\em i.e.} one attempts
to accommodate small black holes, one must take $\epsilon<\sqrt 3-1$ to achieve 
(\ref{newc}) because $N$ is not a large number but $>1$ for a nontrivial Chern-Simons theory.

Na\"{\i}vely, the inequality (\ref{newc}) implies that each $|m|<k/2$; however, the most 
stringent upper bound comes from the configuration with minimum $N$, that is $N=2$. 
In view of the fact that the two $m$ must sum to zero, it follows that the upper bound is 
$|m|<k/4$, which also holds for cases with more than two punctures. 

Now given a flux quantum number $b$ in general there are many spin quantum numbers 
$m=-b/2+nk/2$ where $n\in\mb N$ is some integer which could be zero. 
At most one of these $m$ can be less than $k/4$ in magnitude. To see this, suppose the minimum of 
$|-b/2+nk/2|$ is at $n=n_0$ and $|-b/2+n_0k/2|<k/4$. Then for integer 
$p,|p|\geqslant 1$ the other values $n_0+p$ will not satisfy the bound since 
$|-b/2+(n_0+p)k/2| \geqslant|p|k/2-|-b/2+n_0k/2|>(2|p|-1)k/4\geqslant k/4$. 
But this proof relies on the assumption that there exists an integer $n_0$ for which
$|-b/2+n_0k/2|<k/4$. This assumption breaks down for the special values $b=\pm k/2$, 
in which case the condition requires $|2n_0\pm 1|<1$ for which there is {\em no} 
solution for $n_0$, implying that such a $b$ has no corresponding $m$.

Thus, only if the domain of the surface states is restricted to exclude
the values $b=\pm k/2$ for flux quantum numbers does the map from flux
quantum numbers to the associated spin quantum numbers become one-to-one. 
Now if one takes the point of view that only 
surface states represent true black hole microstates, then one can use this one-to-one
map to count the bulk states carrying only the $m$ quantum numbers.
As the bulk states are characterized by both $j,m$, clearly one must consider 
as equivalent the states that have different $j$ but same $m$.
Note that since the constraints (\ref{arear}) and (\ref{spinc}) 
involve the bulk quantum numbers $j,m$, it is essential to consider the bulk
states in order to take care of the constraints. In fact, the above reasoning
illustrates the fact that the equivalence of $b$ and $m$ quantum numbers has 
nothing to do with how one characterizes states in the effective theory, 
{\em i.e.} whether or not one should regard $j$ quantum numbers as relevant. 
This leaves us at this stage with the choices of $(j,m)$ and $m$-counting.
We shall now argue in favour of one.

%\section {\it Remarks on $(j,m)$ counting:} 
In general, a complete list of quantum numbers to label the states of a black hole 
or of an isolated horizon remains an open issue in loop quantum gravity 
because a complete set of observables is yet to be constructed in the full theory. 
In fact, we expect a full quantum theory of black holes not to be a theory
on the surface alone (unless some sort of holography is at work, which is quite
unlikely in loop quantum gravity). In an effective case, one assumes that only 
a partial set of observables is relevant -- the others being `slow' and not
so relevant for the purpose of describing static or equilibrium properties of
a black hole, contributing only in its dynamical properties -- namely the
area and the flux. While the former is associated with the bulk
Hilbert space, the latter is associated with the surface Hilbert space. 
Boundary conditions provide a one-to-one mapping between the bulk quantum number 
$m$ and the flux quantum number $b$ in the sense explained above. 
The additional $j$ labels, which determine the area, are irrelevant so far as 
the surface states are concerned and this is the reason why the $m$-scheme 
was conceived, but the idea that the Chern-Simons theory represents the effective
theory of the horizon is misleading. An effective theory 
of the horizon is not necessarily a theory {\em on} the horizon. A quantum horizon 
is characterized by a set of flux quantum numbers, an area eigenvalue and a number of boundary
conditions. Only the first one of these is implementable on the surface states alone, 
hence any surface theory is inadequate to define a quantum isolated horizon. 
It is clear that an effective theory of the horizon must involve both a surface 
theory and a bulk theory. 
The bulk degrees of freedom provide -- at least in the description 
that is at hand \cite{ash} -- a relevant observable for the horizon, the area, 
which is kept {\em fixed}. 
Different eigenstates must be used as
physically distinct quantum states and thus the $(j,m)$-scheme is needed. In our
view a genuine field theory of an isolated horizon must be described by the Hilbert 
space ${\cal H}_v\otimes{\cal H}_s$ and hence the surface labels are inadequate to 
characterize a quantum isolated horizon.

%\section
{\it Illustration of state counting:} 
To illustrate the details of the precise counting procedures, we
consider a small black hole with $A=4\sqrt{6}\approx 9.80$.
The {\it exact} eigenvalue corresponds to 2 punctures with $j=2,2$. Each puncture
in principle has 5 allowed values for $m$, but not all the 25
states obey (\ref{newc}), which is satisfied only if
$m=\pm 2,\mp 2$, $m=\pm 1,\mp 1$ or $m=0,0$, so that there are
5 states satisfying those
conditions. However, as indicated above, the $m$ quantum number
has to be non-zero, so there are only 4 states corresponding to
this area in $m$ counting. This is also the number of states in
$(j,m)$ counting.

Instead of fixing the exact eigenvalue at $A=4\sqrt{6}$, we may fix the
classical area at $k=10$ and count states with nearby eigenvalues.
Possibilities with 2 punctures are $j=2,2$($m=\pm2,\mp2$ or $m=\pm1,\mp1$),
$j=\frac52,\frac32$($m=\pm\frac32,\mp\frac32$ or $m=\pm\frac12,\mp\frac12$),
$j=3,1$($m=\pm1,\mp1$) and $j=\frac72,\frac12$($m=\pm\frac12,\mp\frac12$).

Possibilities with 3 punctures are
$j=3,\frac12,\frac12$($m=\pm1,\mp\frac12,\mp\frac12$),
$j=\frac52,1,\frac12$($m=\pm\frac32,\mp1,\mp\frac12$ or $m=\pm\frac12,\mp1,\pm\frac12$),
$j=2,1,1$($m=\pm2,\mp1,\mp1$) and
$j=2,\frac32,\frac12$($m=\pm2,\mp\frac32,\mp\frac12$ or $m=\pm1,\mp\frac32,\pm\frac12$).
There are other possibilities as well.

In $m$ counting,
$j=2,2$($m=\pm1,\mp1$) and $j=3,1$($m=\pm1,\mp1$) are not distinguished,
just as $j=\frac52,\frac32$($m=\pm\frac12,\mp\frac12$) and
$j=\frac72,\frac12$($m=\pm\frac12,\mp\frac12$) are not. However, these
states correspond to distinct areas and $(j,m)$ counting recognizes them
as different. Similarly
$j=3,\frac12,\frac12$($m=\pm1,\mp\frac12,\mp\frac12$) and
$j=1,\frac52,\frac12$($m=\pm1,\mp\frac12,\mp\frac12$) have different
areas but are treated as same in $m$ counting, just as
$j=\frac52,1,\frac12$($m=\pm\frac32,\mp1,\mp\frac12$) and
$j=\frac32,2,\frac12$($m=\pm\frac32,\mp1,\mp\frac12$) are.
Thus the number of states is less in this prescription.

It may be noted that each $|m|\leqslant 2<\frac52$ here, as
expected for this case.

%\section
{\it An  $|m|=j$ rule?} 
Very few of the above states in $m$ counting satisfy the $|m|=j$ rule.
To understand the motivation for setting $|m|=j$,
we may recall a theorem from \cite{dom}.
In $m$ counting, the number of states with  $2\sum\sqrt{j(j+1)}\leqslant A$
is equal to the number of states with $2\sum\sqrt{|m|(|m|+1)}\leqslant A$.
This is useful in the counting of states for large black holes,
where one can consider the range $2\sum\sqrt{j(j+1)}\leqslant A$ instead of
$ A-\epsilon \leqslant 2\sum\sqrt{j(j+1)}\leqslant  A+\epsilon.$
Then one can count {\it only the states with} $|m|=j$. But this can be
done {\it only if large areas are considered}, otherwise the inequality 
$2\sum\sqrt{j(j+1)}\leqslant A$ is not appropriate.

%\section
{\it Counting with SU(2):} 
Let us now discuss the suggested use of $SU(2)$ Chern-Simons theory as the effective 
quantum field theory of isolated horizons \cite{perez}. Two representations 
of $SU(2)$ are involved here; one is associated with the bulk Hilbert space and the 
other with the surface Hilbert space. The surface spin quantum numbers 
$(j_s,m_s)$ will be related to the bulk labels $(j,m)$ via the boundary conditions 
only. In this case the holonomy matrices $h\in SU(2)$ around each puncture 
acting on the surface states is to be matched with the exponentiated triad 
$e$ restricted to the edge that is attached to the 
puncture acting on the bulk states. Since both the operators are $SU(2)$ group 
elements, the matching condition $h\Psi_v\otimes\mb I\Psi_s=\mb I\Psi_v\otimes e\Psi_s$ 
can be implemented strongly only when both $\Psi_s,\Psi_v$ are respectively eigenstates 
of $h,e$ with equal eigenvalues. The result will be a condition like
$m_s=-m$ for each puncture, similar to (\ref{boundc}), together with $j_s=j$, which arises
because the dimensionalities of the bulk and the surface representations have to match.
However, there is another consistency condition on the surface states: the product of all 
holonomies must be the identity, which amounts to the singlet condition that the 
total spin $\sum{\bf j}_s$ vanishes on states. As $j_s=j$, one can rewrite the area 
eigenvalue in terms of surface labels, unlike the $U(1)$ case. 
The singlet condition even takes care of the $\sum m_s=0$ condition more strongly. 
One can count $SU(2)$ Chern-Simons states taking care of these constraints.
The calculation is similar to the $(j,m)$ procedure in the $U(1)$ theory, and
most of the above relations hold here, with minor changes. One difference is
that $m=0$ is not to be excluded here, but only $j=0$, because only the latter
implies a trivial puncture. Another difference is in the relation between the
level $k$ and the classical area, which here is
\beq A_{class}=\frac{1-\gamma^2}{2}k. \eeq
With the appropriate factor in front of $k$, the earlier restrictions on $m$ hold 
once again. Of course, the number of states will be reduced here
by the $SU(2)$ singlet condition.

As an illustration, we may consider the exact area eigenvalue
$A=4\sqrt{6}$ again, so that there are two punctures with
$j=2$ on each. There is only one singlet state that can be
constructed out of two such spins.

%\section{Conclusion}
{\it Conclusion:} 
In summary, if one chooses to count $U(1)$ states with
distinct $m$-labels \cite{dom} instead of states with distinct $j,m$-labels
\cite{gm}, one fails to distinguish between some states
which have different areas. 
The mapping between the allowed values of $b$ and $m$ is one-to-one,
but not all values of $b$ are allowed. 
As regards the $|m|=j$ prescription, it does not even 
count all states with distinct $m$-labels: it
gives a reduced value except in special cases involving $j=\frac12$
or for large area \cite{meissner,gm2}. The $SU(2)$ formulation
is analogous to the $j,m$ procedure and simpler as far as counting is concerned.

\end{document}